\documentstyle[prb,aps,multicol,epsf]{revtex}

\newlength{\colwidth}
\begin{document}
\draft
\title{Electron Refrigeration in the Tunneling Approach}
\author{Heinz--Olaf M{\"u}ller\cite{email} and K. A. Chao}
\address{Department of Physics, Norwegian Institute of Technology,
Norwegian University of Science and Technology,\\
 N-7034 Trondheim, Norway}
\date{\today}

\maketitle
\begin{abstract}
  The qualities of electron refrigeration by means of tunnel junctions
  between superconducting and normal--metal electrodes are studied
  theoretically. A suitable approximation of the basic expression for
  the heat current across those tunnel junctions allows the
  investigation of several features of the device such as its optimal
  bias voltage, its maximal heat current, its optimal working point,
  and the maximally gained temperature reduction. Fortunately, the
  obtained results can be compared with those of a recent experiment.
\end{abstract}

\ifpreprintsty\relax\else\begin{multicols}{2}\fi

\section{Introduction}
\label{intro}

The properties of the electron refrigerator\cite{nah1} are considered
in this paper. In conjunction with the SET--thermometer\cite{pek1} it
forms a new device\cite{lei1} for refrigerating and monitoring
electrons at sub--Kelvin temperature. A future on--chip instrument can
be used to provide isothermal electrons to other experiments, e.g.\ in
single electronics. The operation is similar to a household
refrigerator in so far as the electron temperature, which decouples
from the that of the phonons, can be chosen beforehand within a
certain domain.  Whereas the SET--thermometer, based on ``orthodox
theory'',\cite{ave1} is rather transparent, less effort is taken in
case of the refrigerator.\cite{bar7}

The set-up of the experiment under consideration\cite{lei1} is shown
schematically in Fig.~\ref{fig1}. It consists of a pair of tunnel
junctions forming the refrigerator and another pair operating as
thermometer. All junctions are connected to the island in the middle
whose electrons are dealt with. The leads of the refrigerator are
superconducting whereas the thermometer leads are made from a normal
metal. The refrigerator double junction is biased by a voltage $V$
close to $\Delta/e$, where $\Delta$ is the superconducting gap of its
leads.  This allows only hot electrons (with an energy larger than the
Fermi energy) to leave the island and, vice versa, only cold ones to
enter it via the other junction. It is this dispersion which causes
electron refrigeration at sub--Kelvin temperature, where the heating
by phonons is weak.

On the other hand, the thermometer represents an ultrasmall
single--electron double tunnel junction. The thermal dependence of its
Coulomb blockade enables the determination of the electron
temperature. In general the thermometer is biased too. Since the
zero--bias conductance is sufficient\cite{pek1} and advantageous in
order to determine the temperature, the bias is usually small and so
is the thermometer current as well as the heating by this current.

The tunneling regime of the refrigerator is considered in this paper,
which is in contrast to Ref.~\onlinecite{bar7}. Due to the large
junction resistance Andreev reflections are suppressed and therefore
neglected. Following the argument from above, the heating by the
thermometer current is also neglected. Furthermore, the refrigerating
junctions are thought to be symmetric. This means that there is an
equal voltage drop at each junction and the continuity equation for
the electric current is fulfilled. In the experiment,\cite{lei1}
however, a slight asymmetry was found, which is neglected henceforth.
Additionally, thermalized electrons are supposed, i.e.\ a sufficiently
high temperature causes a sufficiently short electron--electron
scattering time, which in turn causes thermalization of the electrons.
As pointed out in Ref.~\onlinecite{nah1} the electron temperature
should be above $100\,{\rm mK}$ to qualify this assumption. An
opposite point of view,\cite{osw1} however, reveals interesting
physics as well.

Our approach starts from the classical formula for the heat current
across a NIS junction, namely\cite{nah1}
\begin{equation}
  \label{int}
  P(V) = \frac{1}{e^2R}\int{\rm d}\varepsilon N(\varepsilon)
  (\varepsilon-e\,V)[{\rm f}_{\rm N}(\varepsilon-e\,V)-{\rm f}_{\rm
    S}(\varepsilon)],
\end{equation}
where $N(\varepsilon)$ is the density of states on the superconducting
side, which is assumed to be the BCS density of states,
$N(\varepsilon) = |\varepsilon|\Theta(|\varepsilon|-\Delta)/
\sqrt{\varepsilon^2-\Delta^2}$, and ${\rm f}_{\rm
  N,S}(\varepsilon)=1/[\exp(\varepsilon/k_{\rm B}T_{\rm N,S})+1]$ is
the Fermi distribution function for the normal-- and superconducting
side, respectively. $R$ is the normal--state junction resistance. The
integral (\ref{int}) is solved in an analytic approximation in
Sec.~\ref{calc}. Later on conclusions of this result are discussed in
Sec.~\ref{disc}. It might arise the question, what the advantage of
this approximative treatment is if a numerical solution is obtained
readily. In our opinion those formulae yield a better insight into the
behavior of the refrigerator and its qualities and therefore they are
thoroughly worthwhile.

\section{Calculation}
\label{calc}

In term of the BCS density of states $N(\varepsilon)$ Eq.~\ref{int}
can be transformed into
\begin{eqnarray}
  \label{int2}
  P(V) & = & \frac{1}{e^2R}\int_{\Delta}^{\infty}\!\!\!
  {\rm d}\varepsilon\frac{\varepsilon^2}{\sqrt{\varepsilon^2
      -\Delta^2}}\nonumber\\
  & & \times[{\rm f}_{\rm N}(\varepsilon-e\,V)+{\rm f}_{\rm N}
  (\varepsilon+e\,V)-2{\rm f}_{\rm S}(\varepsilon)]\nonumber\\
  & & -\frac{e\,V}{e^2R}\int_{\Delta}^{\infty}\!\!\!
  {\rm d}\varepsilon\frac{\varepsilon}{\sqrt{\varepsilon^2
      -\Delta^2}}\nonumber\\
  & & \times[{\rm f}_{\rm N}(\varepsilon-e\,V)-{\rm f}_{\rm N}
  (\varepsilon+e\,V)].
\end{eqnarray}
In order to simplify the calculation we apply the following
approximation to the Fermi function\cite{kre11}
\begin{equation}
  \label{approx}
  {\rm f}(\varepsilon) \approx \Theta(-\varepsilon)
  +\frac{{\rm sgn}(\varepsilon)}{2}\,{\rm e}^{-\gamma|\varepsilon|}
\end{equation}
with $\gamma=1/(2k_{\rm B}T\ln2)$ and ${\rm sgn}(\varepsilon) =
\varepsilon/|\varepsilon|$, ${\rm sgn}(0)=0$. The value of $\gamma$ is
adjusted to meet
\[
  \int_0^{\infty}\!\!\!{\rm d}\varepsilon\,{\rm f}(\varepsilon)
  = \frac{1}{2}\int_0^{\infty}\!\!\!{\rm d}\varepsilon\,
  {\rm e}^{-\gamma\,\varepsilon}.
\]
In the limit $T\to0$ the approximation is exact. Furthermore, it is
right at $\varepsilon=0$. For large $|\varepsilon|\gg k_{\rm B}T$ the
difference to the exact formula is negligible, but for an intermediate
energy, $\varepsilon\sim k_{\rm B}T$ it is not. Hence, we expect this
approximation to work out at low temperature, i.e.\ at $k_{\rm
  B}T\ll\Delta$ here. This condition is obeyed fairly good in the
experiment.\cite{lei1}

The main advantage of (\ref{approx}), however, is the possibility to
reduce the integrals of (\ref{int2}) to Laplace transformations. In
detail we find with
\begin{eqnarray*}
  {\rm f}_{\rm N}(\varepsilon-e\,V)+{\rm f}_{\rm N}(\varepsilon+e\,V)
  -2{\rm f}_{\rm S}(\varepsilon) & \approx & 
  {\rm e}^{-\gamma_{\rm N}\varepsilon}\cosh(\gamma_{\rm N}e\,V)\\
  & & -{\rm e}^{-\gamma_{\rm S}\varepsilon},\\
  {\rm f}_{\rm N}(\varepsilon-e\,V)-{\rm f}_{\rm N}(\varepsilon+e\,V)
  & \approx & {\rm e}^{-\gamma_{\rm N}\varepsilon}
  \sinh(\gamma_{\rm N}e\,V)
\end{eqnarray*}
and\cite{obe1}
\begin{eqnarray*}
  g(p) = {\cal L}[f(t)](p) & = & \int_0^{\infty}\!\!\!
  {\rm d}tf(t){\rm e}^{-p\,t},\\
  {\cal L}\bigg[\frac{\varepsilon\,\Theta(\varepsilon-\Delta)}
  {\sqrt{\varepsilon^2-\Delta^2}}\bigg](\gamma) & = & \Delta
  {\rm K}_1(\gamma\,\Delta),\\
  {\cal L}\bigg[\varepsilon\frac{\varepsilon\,\Theta(\varepsilon
    -\Delta)}{\sqrt{\varepsilon^2-\Delta^2}}\bigg](\gamma) & = & 
  -\Delta\frac{{\rm d}}{{\rm d}\gamma}{\rm K}_1(\gamma\,\Delta)\\
  & = & \frac{\Delta^2}{2}[{\rm K}_0(\gamma\,\Delta)+{\rm K}_2
  (\gamma\,\Delta)],
\end{eqnarray*}
where ${\rm K}_{\nu}(z)$ is the modified Bessel function of $\nu$--th
order,
\begin{eqnarray}
  \label{pv}
  P(V) & \approx & \frac{\Delta^2}{2e^2R}\big\{[
  {\rm K}_0(\gamma_{\rm N}\Delta)+{\rm K}_2(\gamma_{\rm N}\Delta)]
  \cosh(\gamma_{\rm N}e\,V)\nonumber\\
  & & -{\rm K}_0(\gamma_{\rm S}\Delta)-{\rm K}_2(\gamma_{\rm S}
  \Delta)\big\}\nonumber\\
  & & -\frac{\Delta e\,V\sinh(\gamma_{\rm N}e\,V)}{e^2R}
  {\rm K}_1(\gamma_{\rm N}\Delta).
\end{eqnarray}
Eq.~\ref{pv} is the main result of this paper. We are going to discuss
it in the next section.

\section{Discussion}
\label{disc}

Following Ref.~\onlinecite{bar7} we consider the limit $\gamma_{\rm
  N,S}\Delta\gg1$, $e\,V=\Delta$ first, which describes the experimental
situation to a large extent. In this case the use of the asymptotic
expansion\cite{abr1}
\[
  {\rm K}_{\nu}(z) \approx \sqrt{\frac{\pi}{2\,z}}{\rm e}^{-z}
  \bigg[1+\frac{4\nu^2-1}{8\,z}\bigg]
\]
simplifies (\ref{pv}) to
\begin{eqnarray}
  \label{pv2}
  P(\Delta/e) & \approx & \frac{(\pi\ln2)^{3/2}}{2\,\pi}\,
  \frac{\Delta^2}{e^2R}\bigg(\frac{k_{\rm B}T_{\rm N}}{\Delta}
  \bigg)^{3/2}\nonumber\\
  & \approx & 0.51\,\frac{\Delta^2}{e^2R}
  \bigg(\frac{k_{\rm B}T_{\rm N}}{\Delta}\bigg)^{3/2}.
\end{eqnarray}
Eq.~\ref{pv2} recovers the temperature dependence $P(\Delta/e)\propto
T_{\rm N}^{3/2}$ of the exact result.\cite{bar7} The numerical
prefactor, however, is found to be slightly larger ($0.48$ in
Ref.~\onlinecite{bar7}). The agreement approves of the
introduced approximation (\ref{approx}) and indicates that its
accuracy is in the few--percent range.

In a next step (\ref{pv}) can be used to determine the optimal value
of the bias voltage $V$. From the formal derivative of (\ref{pv}) it
is found that this value $V_{\rm opt}$ should obey the equation
\begin{equation}
  \label{vopt}
  \gamma_{\rm N}e\,V_{\rm opt}\coth(\gamma_{\rm N}e\,V_{\rm opt}) 
  = -\gamma_{\rm N}\Delta\,\frac{{\rm d}\,\ln[{\rm K}_1(\gamma_{\rm N}
    \Delta)]}{{\rm d}(\gamma_{\rm N}\Delta)}-1.
\end{equation}
We like to stress that this result is independent of the lead
temperature $T_{\rm S}$. In the limit $\gamma_{\rm N}\Delta\gg1$ and
$e\,V\sim\Delta$ Eq.~\ref{vopt} simplifies again considerably to
\begin{equation}
  \label{vopt2}
  e\,V_{\rm opt} = \Delta-k_{\rm B}T_{\rm N}\ln2
  \approx \Delta-0.69\,k_{\rm B}T_{\rm N}.
\end{equation}
In Fig.~\ref{fig2} we plot both the numerical solution of (\ref{vopt})
and its approximation (\ref{vopt2}). Up to $k_{\rm B}T_{\rm N}\approx
0.3\Delta$ the approximation works nicely thus covering the relevant
temperature range (see next paragraph). It is difficult to observe
this shift in the experimental data,\cite{lei1} since it corresponds
to maximal $28\,\mu{\rm V}$ only.

In Fig.~\ref{fig3} the heat current (\ref{pv}) using the optimal bias
voltage $V_{\rm opt}$ of (\ref{vopt}) is shown in comparison with both
experimental data and the simpler approximation (\ref{pv2}). The
maximum of the heat current is found at $k_{\rm B}T_{\rm
  N}\le0.23\Delta$ always. Therefore (\ref{vopt2}) is a practical
approximation under relevant conditions as outlined above.  Whereas
Eq.~\ref{pv2} yields a useful rule of thumb, the numerical solution is
rather convincing. Both the optimal temperature and the maximal heat
current are determined satisfactory for higher temperature. For lower
temperature, however, we observe an increasing deviation. This
deviation can not be due to our approximation (\ref{approx}), which
improves with decreasing temperature. According to
Ref.~\onlinecite{lei1} also the thermometer as such is not responsible
for those discrepancies. We attribute the deviation to rising
nonequilibrium effects,\cite{nah1} i.e.\ a non--Fermi like energy
distribution of the electrons on the island. In Ref.~\onlinecite{nah1}
a value of $100\,{\rm mK}$ is given as a lower limit of the
equilibrium description. The exact value, however, depends on the
specific sample layout. It is reasonable to expect an increased limit
for the experiment under consideration since the temperature reduction
is considerably larger and consequently the nonequilibrium effect
becomes more pronounced.

Finally, we are going to discuss an estimate of the lower limit of the
electron temperature $T_{\rm N}$ based on (\ref{pv}). It is assumed
that the heating of the electrons on the island is only due to phonon
coupling\cite{lei1} and follows\cite{wel1}
\begin{equation}
  \label{elph}
  P_{\rm el-ph} = \Sigma\Omega(T_{\rm S}^5-T_{\rm N}^5)
\end{equation}
with the island volume $\Omega$ and a material dependent prefactor
$\Sigma$. Eq.~\ref{elph} implies the assumption that the electrons in
the leads are in thermal equilibrium with the phonons and the phonon
temperature is constant throughout the whole device. For the stationary
situation at optimal bias, $P(V_{\rm opt})=P_{\rm el-ph}$, follows
a relation between $T_{\rm N}$ and $T_{\rm S}$ in terms of
Eqs.~\ref{pv},~\ref{vopt}, and~\ref{elph}. Simpler is the treatment
using (\ref{pv2}) and (\ref{vopt2}) instead of (\ref{pv}) and
(\ref{vopt}), respectively, which results in the
expression
\begin{equation}
  \label{tmin2}
  \bigg(\frac{k_{\rm B}T_{\rm S}}{\Delta}\bigg)^5 = 
  \bigg(\frac{k_{\rm B}T_{\rm N}}{\Delta}\bigg)^5
    +A\,\bigg(\frac{k_{\rm B}T_{\rm N}}{\Delta}\bigg)^{3/2}
\end{equation}
with $A=(\pi\ln2)^{3/2}/(2\pi)\Delta^2/(e^2R\Sigma\Omega) (k_{\rm
  B}/\Delta)^5$. For the maximum temperature reduction, $T_{\rm
  S}-T_{\rm N}$, (\ref{tmin2}) yields
\begin{equation}
  \label{tmin3}
  \bigg(\frac{k_{\rm B}T_{\rm S,opt}}{\Delta}\bigg)^4 = 
  \bigg(\frac{k_{\rm B}T_{\rm N,opt}}{\Delta}\bigg)^4
    +0.3\,A\,\bigg(\frac{k_{\rm B}T_{\rm N,opt}}{\Delta}\bigg)^{1/2}
\end{equation}
for the optimal values of $T_{\rm N,S}$. Using a further approximation,
$T_{\rm N,opt}\ll T_{\rm S,opt}$, we find from (\ref{tmin2}) and
(\ref{tmin3})
\begin{eqnarray}
  \label{topt}
  T_{\rm N,opt} & \approx & 0.3^{10/7}A^{2/7} 
  \approx 0.179\,A^{2/7},\nonumber\\
  T_{\rm S,opt} & \approx & 0.3^{3/7}A^{2/7}\:\: \approx 0.597\,A^{2/7}.
\end{eqnarray}
The last approximation reveals an interesting uniform relation $T_{\rm
  N,opt}/T_{\rm S,opt}\approx0.3$ for the optimal working point of the
refrigerator, independent of all device parameters. Indeed, the
optimal values of the experiment, $T_{\rm N,opt}\approx100\,{\rm mK}$
and $T_{\rm S,opt}\approx300\,{\rm mK}$, almost meet this universal
ratio. Furthermore, since the figure of the ratio assembles from the
general exponents 3/2 and 5 only, it is expected to be right beyond
the range of our approximation.

In Fig.~\ref{fig4} the numerical solution of $P(V_{\rm opt})=P_{\rm
  el-ph}$ in terms of Eqs.~\ref{pv},~\ref{vopt}, and~\ref{elph} is
compared with the approximation (\ref{tmin2}) and the experimental
data. Similarly to the discussion above we attribute the deviations at
low temperature to the manifestation of nonequilibrium. Otherwise we
notice fair agreement of the numerical solution and its approximation
as long as the temperature is not to high. In comparison with the
experiment we compute $T_{\rm N,opt}\approx75\,{\rm mK}$ and $T_{\rm
  S,opt}\approx250\,{\rm mK}$ from (\ref{topt}). This values are
somewhat lower than the measured data (see last paragraph). 

% In addition to the nonequilibrium, cited above, these discrepancies
% probably reveal the limits of the used approximation as well.

\section{Conclusion}

In this paper electron refrigeration by means of tunnel junctions is
studied theoretically. An approximation of the original expression of
the heat current across those junctions enables a closed analytic
formula for this quantity. Based on this formula several features of
the refrigerator are discussed. The optimal bias voltage is determined
and the proportionality of the corresponding heat current to $T_{\rm
  N}^{3/2}$ is confirmed. The maximal difference of the electron
temperature in the leads, $T_{\rm S}$, and the island, $T_{\rm N}$, is
discussed in stationary regime. Analytic approximations
to the optimal values of these temperatures are derived and an
universal ratio $T_{\rm N,opt}/T_{\rm S,opt}=0.3$ is found. All
results are discussed in comparison with a recent experiment. The
investigation of nonequilibrium effects, that occur at very low
temperature, remains an interesting open problem.

\section*{Acknowledgments}

We are indebted to M. M. Leivo and J. P. Pekola for discussion on this
subject and for leaving their paper to us prior to publication. One
of us (HOM) gratefully acknowledges financial support by Deutscher
Akademischer Austauschdienst.

% \bibliography{set}

\begin{references}

\bibitem{email}
hom@phys.unit.no.

\bibitem{nah1}
M. Nahum, T.~M. Eiles, and J.~M. Martinis, Appl. Phys. Lett. {\bf 65},  3123
  (1994).

\bibitem{pek1}
J. Pekola, K. Hirvi, J. Kauppinen, and M. Paalanen, Phys. Rev. Lett. {\bf 73},
  2903  (1994).

\bibitem{lei1}
M.~M. Leivo, J.~P. Pekola, and D.~V. Averin, Appl. Phys. Lett. {\bf 68},  1996
  (1995), cond--mat/9511127.

\bibitem{ave1}
D.~V. Averin and K.~K. Likharev, J. Low Temp. Phys. {\bf 62},  345  (1986).

\bibitem{bar7}
A. Bardas and D.~V. Averin, Phys. Rev. B {\bf 52},  12873  (1995),
  cond--mat/9505097.

\bibitem{osw1}
J. Oswald, http://nahum--www.harvard.edu/expts/noneq.html (unpublished).

\bibitem{kre11}
W. Krech and H.-O. M{\"u}ller, Mod. Phys. Lett. B {\bf 8},  605  (1994).

\bibitem{obe1}
F. Oberhettinger and L. Badii, {\em Tables of Laplace Transforms} (Springer,
  Berlin, Heidelberg, 1973).

\bibitem{abr1}
M. Abramowitz and I.~A. Stegun, {\em Pocketbook of Mathematical Functions}
  (Harri Deutsch, Thun, 1984).

\bibitem{wel1}
F.~C. Wellstood, C. Urbina, and J. Clarke, Phys. Rev. B {\bf 49},  5942
  (1994).

\end{references}
% \bibliographystyle{prsty}

\narrowtext

\ifx\epsfxsize\undefined\relax\else
\ifx\colwidth\undefined
\epsfxsize=\textwidth
\else
\epsfxsize=\colwidth
\fi
\epsffile{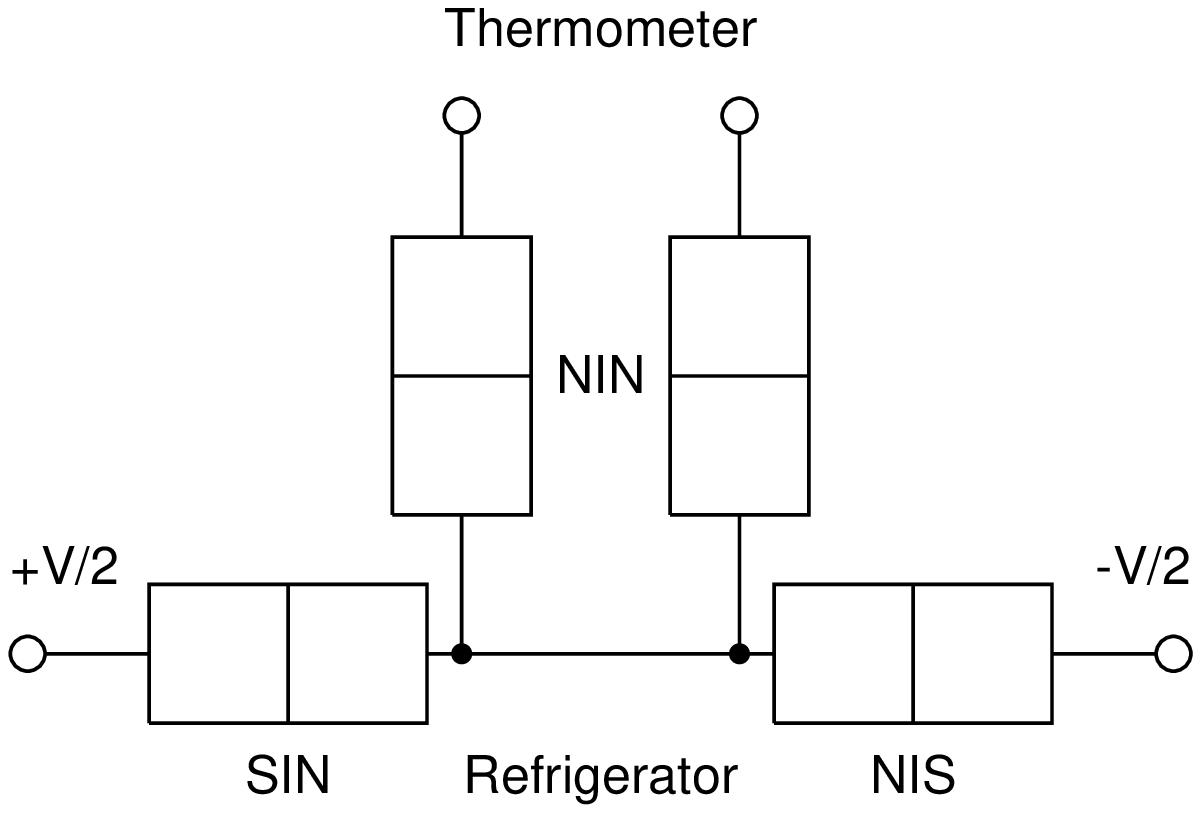}
\fi

\begin{figure}
\caption{Combination of refrigerator and thermometer, both based upon
  tunnel junctions. The horizontally shown refrigerator consists of a
  superconductor--insulator--normal-metal (SIN) junction to the left
  and a NIS--junction to the right biased by a voltage $V$. The
  vertically displayed thermometer tests the electron temperature of
  the island in the middle by means of two NIN--junctions.}
\label{fig1}
\end{figure}

\ifx\epsfxsize\undefined\relax\else
\ifx\colwidth\undefined
\epsfxsize=\textwidth
\else
\epsfxsize=\colwidth
\fi
\epsffile{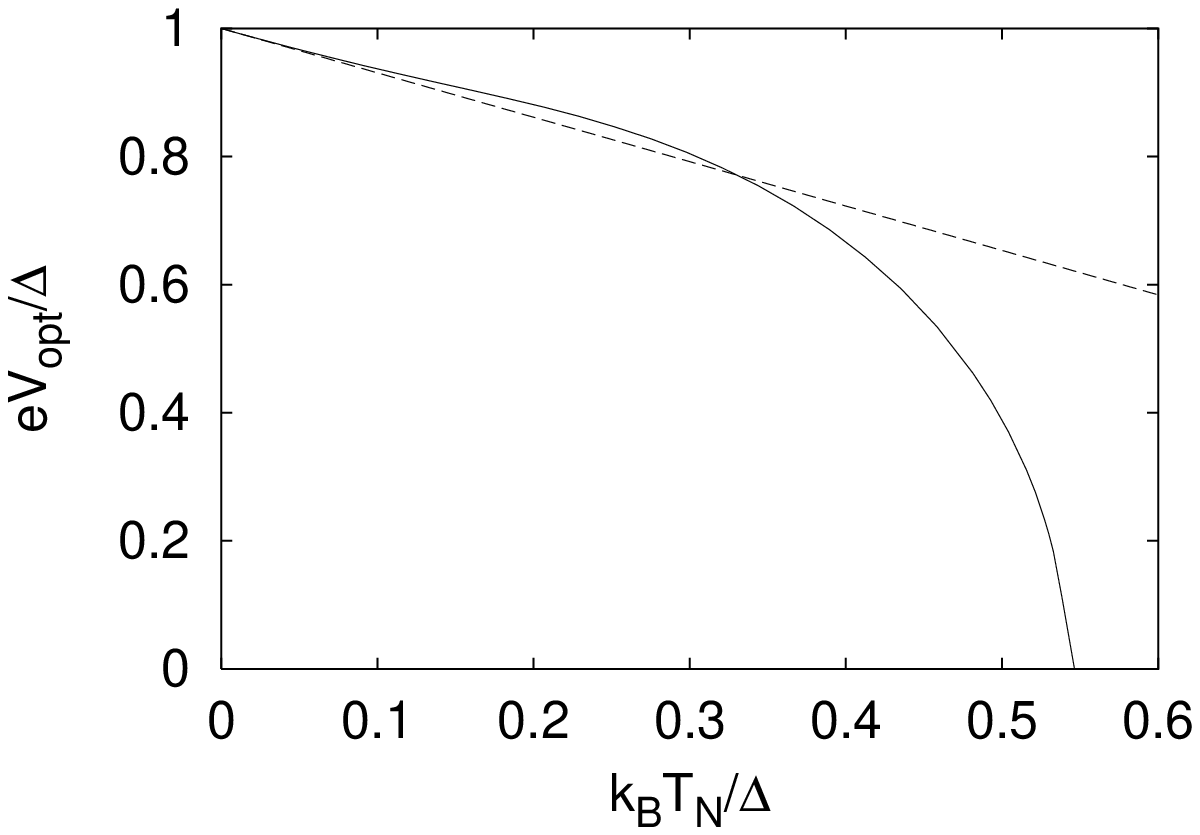}
\fi

\begin{figure}
\caption{Optimal bias voltage $V_{\rm opt}$ in dependence on the
  electron temperature on the island $T_{\rm N}$. The solid curve
  displays the numerical solution of (\ref{vopt}) and the dashed line
  its approximation (\ref{vopt2}).}
\label{fig2}
\end{figure}

\ifx\epsfxsize\undefined\relax\else
\ifx\colwidth\undefined
\epsfxsize=\textwidth
\else
\epsfxsize=\colwidth
\fi
\epsffile{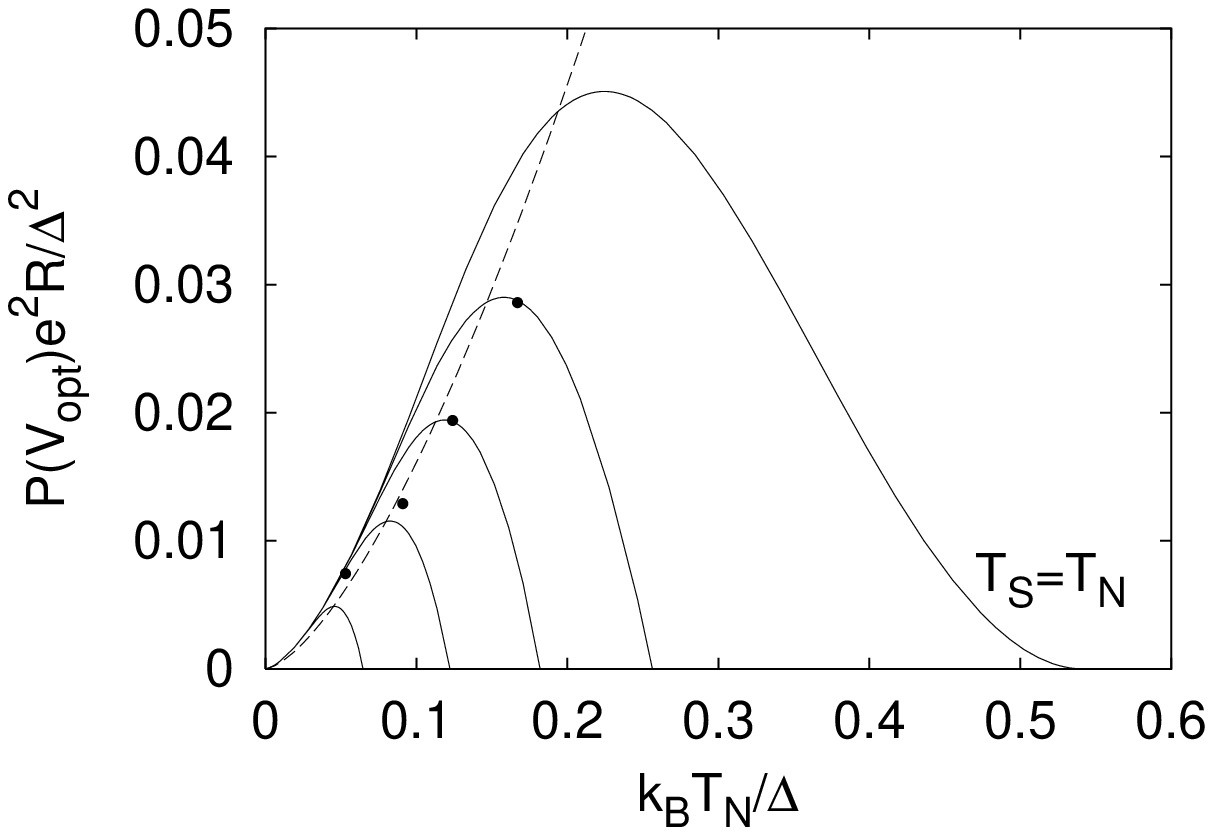}
\fi

\begin{figure}
\caption{The maximal heat current $P(V_{\rm opt})$ as a function of
  the electron temperature $T_{\rm N}$. The solid lines correspond to
  different ratios $T_{\rm S}/T_{\rm N}$ of the electron temperature
  in the leads and the island. For the uppermost curve this ratio is
  $1$, whereas it corresponds to the values of the experiment for the
  other curves. The dashed line shows Eq.~\ref{pv2}, and the dots
  ($\bullet$) correspond to the values of the experiment.}
\label{fig3}
\end{figure}

\ifx\epsfxsize\undefined\relax\else
\ifx\colwidth\undefined
\epsfxsize=\textwidth
\else
\epsfxsize=\colwidth
\fi
\epsffile{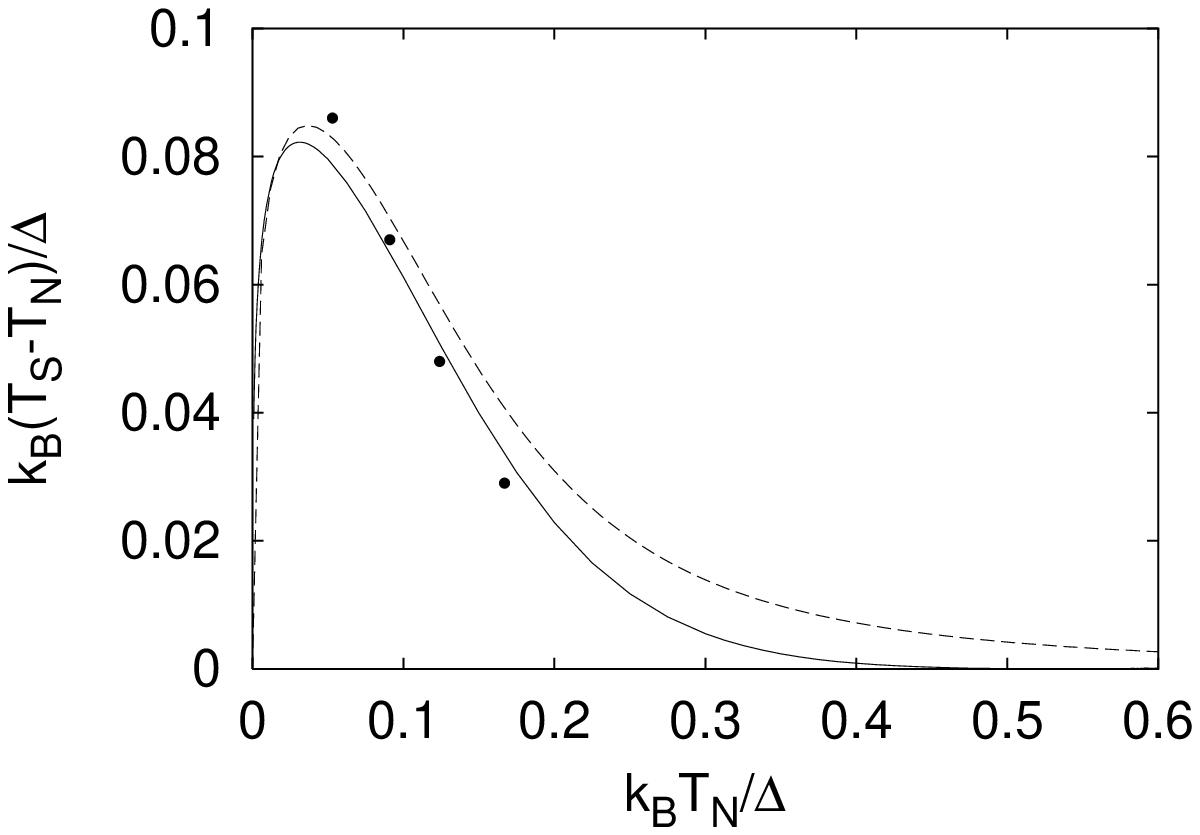}
\fi

\begin{figure}
\caption{Temperature reduction $T_{\rm S}-T_{\rm N}$ versus $T_{\rm
    N}$. The solid curve corresponds to the numerical procedure
  described in the text, whereas the dashed line is a plot of
  (\ref{tmin2}). The symbols ($\bullet$) are again the data of the
  experiment.}
\label{fig4}
\end{figure}

\ifpreprintsty\relax\else\end{multicols}\fi

\end{document}